\documentclass[11pt]{article}
\usepackage{amssymb,amsmath,amsfonts}
\usepackage{graphicx}
\usepackage{graphics}
\usepackage{eepic,epsfig}

1.60cm \makeatletter \@addtoreset{equation}{section}

\makeatother

\begin{document}

\title{Fermionic current induced by magnetic flux \\ in compactified
cosmic string spacetime}
\author{E. R. Bezerra de Mello$^{1}$\thanks{%
E-mail: emello@fisica.ufpb.br},\, A. A. Saharian$^{2}$\thanks{%
E-mail: saharian@ysu.am} \\
\\
\textit{$^{1}$Departamento de F\'{\i}sica, Universidade Federal da Para\'{\i}%
ba}\\
\textit{58.059-970, Caixa Postal 5.008, Jo\~{a}o Pessoa, PB, Brazil}\vspace{%
0.3cm}\\
\textit{$^2$Department of Physics, Yerevan State University,}\\
\textit{1 Alex Manoogian Street, 0025 Yerevan, Armenia}}
\maketitle

\begin{abstract}
In this paper, we investigate the fermionic current densities induced by a
magnetic flux running along the idealized cosmic string in a
four-dimensional spacetime, admitting that the coordinate along the string's
axis is compactified. In order to develop this investigation we construct
the complete set of fermionic mode functions obeying a general
quasiperiodicity condition along the compactified dimension. The vacuum
expectation value of the azimuthal current density is decomposed into two
parts. The first one corresponds to the uncompactified cosmic string
geometry and the second one is the correction induced by the
compactification. For the first part we provide a closed expression which
includes various special cases previously discussed in the literature. The
second part is an odd periodic function of the magnetic flux along the
string axis with the period equal to the flux quantum and it is an even
function of the magnetic flux enclosed by the string axis. The
compactification of the cosmic string axis in combination with the
quasiperiodicity condition leads to the nonzero axial current density. The
latter is an even periodic function of the magnetic flux along the string
axis and an odd periodic function of the magnetic flux enclosed by the
string axis. The axial current density vanishes for untwisted and twisted
fields in the absence of the magnetic flux enclosed by the string axis. The
asymptotic behavior of the vacuum fermionic current is investigated near the
string and at large distances from it. In particular, the topological part
of the azimuthal current and the axial current are finite on the string's
axis.
\end{abstract}

\bigskip

PACS numbers: 98.80.Cq, 11.10.Gh, 11.27.+d

\bigskip

\section{Introduction}

In quantum field theory, because of quantum nature of the vacuum state, its
properties depend crucially on both the local geometry and topology of the
background spacetime. In particular, the nontrivial topology leads to the
change of the vacuum energy. This is the topological Casimir effect which
has been investigated for large number of geometries (see, for instance,
\cite{Most97}). In higher-dimensional models with compact extra dimensions,
the dependence of the vacuum energy on the length scale of the compact
subspace provides a stabilization mechanism for moduli fields. In addition,
the topological Casimir effect can be a source of dark energy driving the
accelerating expansion of the Universe at recent epoch \cite{Eliz01}. In the
present paper we consider an exactly solvable problem for a charged massive
fermionic field with two types of the topological vacuum polarization. The
first one is generated by a planar angle deficit in the cosmic string
geometry and the second one is induced by the compactification of the string
axis. Recently, the calculations of topological Casimir densities associated
with a quantum scalar field in compactified cosmic string spacetime, has
been developed in \cite{Mello12}.

The cosmic strings are among the most interesting topological defects which
may have been created by phase transitions in the early Universe \cite{V-S}.
Though the observation data on the cosmic microwave background have ruled
out cosmic strings as primary source for primordial density perturbations,
they are still candidate for the generation of a number of interesting
physical effects such as gamma ray bursts \cite{Berezinski}, gravitational
waves \cite{Damour} and high energy cosmic rays \cite{Bhattacharjee}.
Recently, cosmic strings have attracted renewed interest, partly because a
variant of their formation mechanism is proposed in the framework of brane
inflation \cite{Sarangi}-\cite{Dvali}.

The geometry of the spacetime produced by an idealized infinite straight
cosmic has a conical structure. It is locally flat except on the top of the
string where it has a delta shaped curvature tensor. This nontrivial
structure raises a number of interesting physical effects. One of these
concerns the effect of a string on the properties of quantum vacuum.
Explicit calculations for vacuum expectation values of the energy-momentum
tensor associated with various fields in the vicinity of a string have been
done in \cite{Hell86}-\cite{BezeKh06}. Moreover, considering the presence of
a magnetic flux running along the cosmic strings, there appear additional
contributions to the corresponding vacuum polarization effects associated
with charged fields \cite{charged}-\cite{Spin1}.\footnote{%
Recently the fluxes by gauge fields play an important role in
higher-dimensional models including braneworld scenarios (see, for example,
\cite{Dou}).} The presence of a magnetic flux along the cosmic string
induces vacuum current densities, as well. This phenomenon has been
investigated for massless scalar field in \cite{Sira} and recently for
massive one in \cite{Yu}. There the authors have shown that the induced
vacuum current densities along the azimuthal direction appear if the ratio
of the magnetic flux by the quantum one has a nonzero fractional part. The
analysis of induced fermionic currents in higher-dimensional cosmic string
spacetime in the presence of a magnetic flux have been developed in \cite%
{Mello10}. Moreover, induced fermionic current by a magnetic flux in $(2+1)$%
-dimensional conical spacetime and in presence of a circular boundary has
also been analyzed in \cite{Saha10} (for combined effects of topology and
boundaries on the quantum vacuum in the geometry of a cosmic string see \cite%
{Brev95}).

Another type of topological quantum effects takes place in models which
present compact spatial dimensions. The presence of compact dimensions is an
important feature of most high-energy theories of fundamental physics,
including supergravity and superstring theories. An interesting application
of the field theoretical models with compact dimensions recently appeared in
nanophysics. The long-wavelength description of the electronic states in
graphene can be formulated in terms of the Dirac-like theory in
three-dimensional spacetime with the Fermi velocity playing the role of
speed of light (see, e.g., \cite{Cast09}). Single-walled carbon nanotubes
are generated by rolling up a graphene sheet to form a cylinder and the
background spacetime for the corresponding Dirac-like theory has topology $%
R^{2}\times S^{1}$. In this paper we shall analyze the influence of
compactification of the spatial dimension along the cosmic string's axis, on
the vacuum expectation value (VEV) of the fermionic current induced by a
magnetic flux running along this axis and by the magnetic flux enclosed by
the string axis. This VEV is among the most important local characteristics
of the quantum vacuum. In addition to describing the physical structure of a
charged quantum field at a given point, the current acts as the source in
the Maxwell equations and plays an important role in modeling a
self-consistent dynamics involving the electromagnetic field.

We have organized the paper as follows. In Section \ref{sec2} we present the
background geometry associated with the spacetime under consideration and
construct the complete set of normalized positive- and negative-energy
fermionic wave-functions obeying quasiperiodic boundary condition with an
arbitrary phase along the string axis. In Section \ref{sec3}, by using the
mode-summation method, we evaluate the renormalized fermionic vacuum current
density induced by a magnetic flux running along the string's axis. As we
shall see, the charge density and the radial current vanish. The azimuthal
current density is decomposed into two parts: the first one corresponding to
the geometry of a cosmic string without compactification and the second one
being induced by the compactification. As a consequence of the string axis
compactification, there appears a non-vanishing axial current. This is a
purely topological effect and is investigated in Section \ref{sec4} for the
general case of periodicity conditions along the compact dimension. In
addition, we assume the presence of a constant gauge field with the
non-vanishing component along the string axis. The most relevant conclusions
of the paper are summarized in Section \ref{conc}. Throughout the paper we
use the units with $G=\hbar =c=1$.

\section{Fermionic wave-functions}

\label{sec2}

The main objective of this section is to obtain the complete set of
normalized fermionic wave-functions in a four-dimensional cosmic string
spacetime considering that these fields obey a quasiperiodicity condition
along the string's axis. This set is needed for the calculation of the VEVs
of the fermionic current densities by using the mode-summation approach.

The background geometry of the spacetime corresponding to a cosmic string
along the $z$-axis, can be given, by using cylindrical coordinates, through
the line element below:
\begin{equation}
ds^{2}=dt^{2}-dr^{2}-r^{2}d\phi ^{2}-dz{}^{2}\ ,  \label{ds21}
\end{equation}%
where the coordinates take values in the ranges $r\geqslant 0$, $0\leqslant
\phi \leqslant \phi _{0}=2\pi /q$, $-\infty <t<+\infty $. The parameter $q$,
which codify the planar angle deficit, is related to the mass per unit
length of the string, $\mu _{0}$, by $q^{-1}=1-4\mu _{0}$. Additionally we
shall assume that the direction along the $z$-axis is compactified to a
circle with the length $L$: $0\leqslant z\leqslant L$.

The quantum dynamic of a massive charged spinor field in curved spacetime
and in the presence of a electromagnetic four-vector potential, $A_{\mu }$,
is governed by the Dirac equation
\begin{equation}
i\gamma ^{\mu }(\nabla _{\mu }+ieA_{\mu })\psi -m\psi =0\ ,\ \nabla _{\mu
}=\partial _{\mu }+\Gamma _{\mu }\ ,  \label{Direq}
\end{equation}%
where $\gamma ^{\mu }$ are the Dirac matrices in curved spacetime and $%
\Gamma _{\mu }$ is the spin connection. Both matrices are given in terms of
the flat spacetime Dirac matrices, $\gamma ^{(a)}$, by the relations,
\begin{equation}
\gamma ^{\mu }=e_{(a)}^{\mu }\gamma ^{(a)}\ ,\ \Gamma _{\mu }=\frac{1}{4}%
\gamma ^{(a)}\gamma ^{(b)}e_{(a)}^{\nu }e_{(b)\nu ;\mu }\ .  \label{Gammamu}
\end{equation}%
In (\ref{Gammamu}), $e_{(a)}^{\mu }$ represents the tetrad basis satisfying
the relation $e_{(a)}^{\mu }e_{(b)}^{\nu }\eta ^{ab}=g^{\mu \nu }$, with $%
\eta ^{ab}$ being the Minkowski spacetime metric tensor. We assume that
along the compact dimension the field obeys the quasiperiodicity condition
\begin{equation}
\psi (t,r,\phi ,z+L)=e^{2\pi i\beta }\psi (t,r,\phi ,z)\ ,  \label{Period}
\end{equation}%
with a constant phase $\beta $, $0\leqslant \beta \leqslant 1$. The special
cases $\beta =0$ and $\beta =1/2$ correspond to the periodic and
antiperiodic boundary conditions (untwisted and twisted fields respectively).

In the discussion below we admit the existence of a constant gauge field
with the vector potential%
\begin{equation}
A_{\mu }=(0,0,A_{\phi },A_{z})\ .  \label{Amu}
\end{equation}%
The component $A_{\phi }$ is related to an infinitesimal thin magnetic flux,
$\Phi _{2}$, running along the string by $A_{\phi }=-q\Phi _{2}/(2\pi )$.
Similarly, the axial component $A_{z}$ can be given in terms of the magnetic
flux $\Phi _{3}$ enclosed by the $z$-axis by the relation $A_{z}=-\Phi _{3}/L
$. Though the magnetic field strength corresponding to (\ref{Amu}) vanishes,
the nontrivial topology of the background geometry leads to
Aharonov-Bohm-like effects on the VEVs of physical observables.

In order to find the complete set of mode functions in the problem under
consideration, we shall use the standard representation of the flat space
Dirac matrices:
\begin{equation}
\gamma ^{(0)}=\left(
\begin{array}{cc}
1 & 0 \\
0 & -1%
\end{array}%
\right) ,\;\gamma ^{(a)}=\left(
\begin{array}{cc}
0 & \sigma _{a} \\
-\sigma _{a} & 0%
\end{array}%
\right) ,\;a=1,2,3\ ,  \label{gam0l}
\end{equation}%
with $\sigma _{1},\sigma _{2},\sigma _{3}$ being the Pauli matrices. We take
the tetrad basis as follows:
\begin{equation}
e_{(a)}^{\mu }=\left(
\begin{array}{cccc}
1 & 0 & 0 & 0 \\
0 & \cos (q\phi ) & -\sin (q\phi )/r & 0 \\
0 & \sin (q\phi ) & \cos (q\phi )/r & 0 \\
0 & 0 & 0 & 1%
\end{array}%
\right) \ ,  \label{emua}
\end{equation}%
where the index $a$ identifies the rows of the matrix. With this choice, the
gamma matrices take the form
\begin{equation}
\gamma ^{0}=\gamma ^{(0)},\;\gamma ^{l}=\left(
\begin{array}{cc}
0 & \sigma ^{l} \\
-\sigma ^{l} & 0%
\end{array}%
\right) \ ,  \label{gamcurved}
\end{equation}%
where we have introduced the $2\times 2$ matrices for $l=(r,\ \phi ,\ z)$:
\begin{equation}
\sigma ^{r}=\left(
\begin{array}{cc}
0 & e^{-iq\phi } \\
e^{iq\phi } & 0%
\end{array}%
\right) \ ,\ \sigma ^{\phi }=-\frac{i}{r}\left(
\begin{array}{cc}
0 & e^{-iq\phi } \\
-e^{iq\phi } & 0%
\end{array}%
\right) \ ,\ \sigma ^{z}=\left(
\begin{array}{cc}
1 & 0 \\
0 & -1%
\end{array}%
\right) \ .  \label{betl}
\end{equation}%
For the spin connection and the combination appearing in the Dirac equation
we find
\begin{equation}
\Gamma _{\mu }=\frac{1-q}{2}\gamma ^{(1)}\gamma ^{(2)}\delta _{\mu }^{\phi
}\ ,\ \gamma ^{\mu }\Gamma _{\mu }=\frac{1-q}{2r}\gamma ^{r},  \label{gammu}
\end{equation}%
and the Dirac equation takes the form%
\begin{equation}
\left( \gamma ^{\mu }\left( \partial _{\mu }+ieA_{\mu }\right) +\frac{1-q}{2r%
}\gamma ^{r}+im\right) \psi =0\ .  \label{Direq1}
\end{equation}

For positive energy solutions, assuming the time-dependence of the
eigenfunctions in the form $e^{-iEt}$ and decomposing the spinor $\psi $
into the upper and lower components, denoted by $\psi _{+}$ and $\psi _{-}$,
respectively, we find the equations%
\begin{eqnarray}
\left( \sigma ^{l}(\partial _{l}+ieA_{l})+\frac{1-q}{2r}\sigma ^{r}\right)
\psi _{+}-i\left( E+m\right) \psi _{-} &=&0\ ,  \notag \\
\left( \sigma ^{l}(\partial _{l}+ieA_{l})+\frac{1-q}{2r}\sigma ^{r}\right)
\psi _{-}-i\left( E-m\right) \psi _{+} &=&0\ .  \label{phixieq}
\end{eqnarray}%
Substituting the function $\psi _{-}$ from the first equation into the
second one, we obtain the second order differential equation for the spinor $%
\psi _{+}$:
\begin{equation}
\left[ \partial _{r}^{2}+\frac{1}{r}\partial _{r}+\frac{1}{r^{2}}\left(
\partial _{\phi }+ieA_{2}-i\frac{1-q}{2}\sigma ^{z}\right) ^{2}+\left(
\partial _{z}+ieA_{3}\right) ^{2}+E^{2}-m^{2}\right] \psi _{+}=0\ .
\label{phieq}
\end{equation}%
The same equation is obtained for the spinor $\psi _{-}$.

In order to look for solution for (\ref{phieq}), we use the the ansatz
below, compatible with the cylindrical symmetry of the physical system:
\begin{equation}
\psi _{+}=e^{ikz}\left(
\begin{array}{c}
C_{1}R_{1}(r)e^{iqn_{1}\phi } \\
C_{2}R_{2}(r)e^{iqn_{2}\phi }%
\end{array}%
\right) \ ,  \label{psi+}
\end{equation}%
with $n_{l}$, for $l=1,2$, being integer numbers and $C_{1}$ and $C_{2}$ are
two arbitrary constants. Substituting this function into (\ref{phieq}), we
can see that the solutions of the equations for the radial functions,
regular on the string, are expressed in terms of the Bessel function of the
first kind: $R_{l}(r)=J_{|\nu _{l}|}(\lambda r)$, where the order is given
below,
\begin{equation}
\nu _{1}=q\left( n_{1}+1/2\right) +eA_{2}-1/2\ ,\ \nu _{2}=q\left(
n_{2}-1/2\right) +eA_{2}+1/2\ ,  \label{nu12}
\end{equation}%
and the variable $\lambda $ is related with the energy by,
\begin{equation}
\lambda =\sqrt{E^{2}-\tilde{k}_{+}^{2}-m^{2}}\ ,  \label{lambda}
\end{equation}%
where $\tilde{k}_{+}=k+eA_{3}$. We may see that $|\nu _{2}|=|\nu
_{1}|+\epsilon _{\nu _{1}}$, where $\epsilon _{\nu _{1}}$ is equal to $+1$
for $\nu _{1}\geq 0$ and $-1$ for $\nu _{1}<0$.

Having the upper two-component spinor, we can find the components of the
lower one, $\psi _{-}$, by using the first equation in (\ref{phixieq}). From
this equation we find the following relations,
\begin{equation}
\psi _{-}=e^{ikz}\left(
\begin{array}{c}
B_{1}R_{1}(r)e^{iqn_{1}\phi } \\
B_{2}R_{2}(r)e^{iqn_{2}\phi }%
\end{array}%
\right) \ ,  \label{psilow}
\end{equation}%
with
\begin{equation}
n_{2}=n_{1}+1\ .
\end{equation}%
The coefficients $B_{1}$ and $B_{2}$ are related with $C_{1}$ and $C_{2}$
by,
\begin{eqnarray}
B_{1} &=&\frac{1}{E+m}\left( C_{1}\tilde{k}_{+}-iC_{2}\epsilon _{\nu
_{1}}\lambda \right) \ ,  \notag \\
B_{2} &=&\frac{1}{E+m}\left( iC_{1}\lambda \epsilon _{\nu _{1}}-C_{2}\tilde{k%
}_{+}\right) \ .  \label{B12}
\end{eqnarray}

Finally we can write the positive energy solution in its complete form:
\begin{equation}
\psi =\left(
\begin{array}{c}
C_{1}\ J_{|\nu _{1}|}(\lambda r) \\
C_{2}J_{|\nu _{2}|}(\lambda r)e^{iq\phi } \\
B_{1}\ J_{|\nu _{1}|}(\lambda r) \\
B_{2}J_{|\nu _{2}|}(\lambda r)e^{iq\phi }%
\end{array}%
\right) \ e^{ikz+iqn\phi -iEt},  \label{psi}
\end{equation}%
where we have defined $n_{1}=n$. We can see that (\ref{psi}) is an
eigenfunction of the total angular momentum along the cosmic string:
\begin{equation}
\widehat{J}_{3}\psi =\left( -i\partial _{\phi }+i\frac{q}{2}\gamma
^{(1)}\gamma ^{(2)}\right) \psi =qj\psi \ ,  \label{J3}
\end{equation}%
where
\begin{equation}
j=n+1/2\ ,\ j=\pm 1/2,\pm 3/2,\ldots \ .  \label{j}
\end{equation}

The fermionic wave-function above contains four coefficients and there are
two equations relating them. The normalization condition on the functions
provides an extra equation. Consequently, one of the coefficients remains
arbitrary. In order to determine this coefficient some additional condition
should be imposed on the coefficients. The necessity for this condition is
related to the fact that the quantum numbers $(\lambda ,k,j)$ do not specify
the fermionic wave-function uniquely and some additional quantum number is
required.

In order to specify the second constant we impose the condition
\begin{equation}
C_{1}/B_{1}=-C_{2}/B_{2}\ .  \label{AdCond}
\end{equation}%
By taking into account (\ref{B12}), we can write,
\begin{equation}
C_{2}=sC_{1}\ ,\ B_{1}=-sB_{2}=\frac{\tilde{k}_{+}-is\epsilon _{\nu
_{1}}\lambda }{E+m}C_{1}\ ,\ s=\pm 1\ .
\end{equation}%
With the condition (\ref{AdCond}), the fermionic mode functions are uniquely
specified by the set of quantum numbers $\sigma =(\lambda ,k,j,s)$. The
eigenvalues of the quantum number $k$ are determined from the periodicity
condition (\ref{Period}):
\begin{equation}
k=k_{l}^{(+)}=2\pi (l+\beta )/L,\;l=0,\pm 1,\pm 2,\ldots \ .  \label{Eigkz}
\end{equation}

On the basis of all these considerations, the positive-energy fermionic wave
function is written in the form

\begin{equation}
\psi _{\sigma }^{(+)}(x)=C_{\sigma }^{(+)}\left(
\begin{array}{c}
J_{\beta _{j}}(\lambda r) \\
sJ_{\beta _{j}+\epsilon _{j}}(\lambda r)e^{iq\phi } \\
\frac{\tilde{k}_{+}-is\epsilon _{j}\lambda }{E_{+}+m}J_{\beta _{j}}(\lambda
r) \\
-s\frac{\tilde{k}_{+}-is\epsilon _{j}\lambda }{E_{+}+m}J_{\beta
_{j}+\epsilon _{j}}(\lambda r)e^{iq\phi }%
\end{array}%
\right) e^{ik_{l}^{(+)}z+iq(j-1/2)\phi -iE_{+}t}\ ,  \label{psi+n}
\end{equation}%
where $\epsilon _{j}=1$ for $j>-\alpha $ and $\epsilon _{j}=-1$ for $%
j<-\alpha $, and
\begin{equation}
\beta _{j}=q|j+\alpha |-\epsilon _{j}/2.  \label{betaj}
\end{equation}%
The energy is expressed in terms of $\lambda $ and $\tilde{k}_{+}$ by the
relation
\begin{equation}
E_{+}=\sqrt{\lambda ^{2}+\tilde{k}_{+}^{2}+m^{2}}\ ,  \label{E+}
\end{equation}%
where $\tilde{k}_{+}=2\pi (l+\tilde{\beta})/L$, with%
\begin{equation}
\tilde{\beta}=\beta +eA_{3}L/(2\pi )=\beta -\Phi _{3}/\Phi _{0},
\label{bett}
\end{equation}%
and with $\Phi _{0}=2\pi /e$ being the flux quantum. In (\ref{betaj}) we
have defined%
\begin{equation}
\alpha =eA_{2}/q=-\Phi _{2}/\Phi _{0},  \label{alfa}
\end{equation}
The constant $C_{\sigma }^{(+)}$ is found from the normalization condition
\begin{equation}
\int d^{3}x\sqrt{\gamma }\ (\psi _{\sigma }^{(+)})^{\dagger }\psi _{\sigma
^{\prime }}^{(+)}=\delta _{\sigma \sigma ^{\prime }}\ ,  \label{normcond}
\end{equation}%
where $\gamma $ is the determinant of the spatial metric tensor. The delta
symbol on the right-hand side is understood as the Dirac delta function for
continuous quantum numbers ($\lambda $) and the Kronecker delta for discrete
ones $(k,j,s)$. From (\ref{normcond}) one finds%
\begin{equation}
|C_{\sigma }^{(+)}|^{2}=\frac{q\lambda (E_{+}+m)}{8\pi LE_{+}}\ .  \label{C+}
\end{equation}

The physical results will depend on the phases in the periodicity condition (%
\ref{Period}) and on the component of the gauge potential along the axis of
the string in the form of the combination (\ref{bett}). This could be seen
directly by noting that the axial component of the vector potential is
excluded from the field equation (\ref{Direq}) by the gauge transformation $%
A_{\mu }^{\prime }=A_{\mu }+\partial _{\mu }\Lambda $, $\psi ^{\prime
}(x)=e^{-ie\Lambda }\psi (x)$ with$\;\Lambda =-A_{3}z$. In the new gauge one
has $A_{3}^{\prime }=0$ and the periodicity condition has the form $\psi
^{\prime }(t,r,\phi ,z+L)=e^{2\pi i\tilde{\beta }}\psi ^{\prime }(t,r,\phi
,z)$. Hence, the presence of the component of the gauge field along compact
dimension is equivalent to the shift in the phase of the corresponding
periodicity condition. In particular, a non-trivial phase is induced for
untwisted fields.

The negative-energy fermionic mode-function can be obtained in a similar
way. The corresponding result is given by the expression:
\begin{equation}
\psi _{\sigma }^{(-)}(x)=C_{\sigma }^{(-)}\left(
\begin{array}{c}
J_{\beta _{j}+\epsilon _{j}}(\lambda r)e^{-iq\phi } \\
sJ_{\beta _{j}}(\lambda r) \\
-\frac{\tilde{k}_{-}-is\epsilon _{j}\lambda }{E_{-}-m}J_{\beta _{j}+\epsilon
_{j}}(\lambda r)e^{-iq\phi } \\
s\frac{\tilde{k}_{-}-is\epsilon _{j}\lambda }{E_{-}-m}J_{\beta _{j}}(\lambda
r)%
\end{array}%
\right) e^{iE_{-}t-iq(j-1/2)\phi -ik_{l}^{(-)}z}\ ,  \label{psi-}
\end{equation}%
with $k_{l}^{(-)}=2\pi (l-\beta )/L$,$\;l=0,\pm 1,\pm 2,\ldots $, $\tilde{k}%
_{-}=2\pi (l-\tilde{\beta})/L$, and%
\begin{equation}
E_{-}=\sqrt{\lambda ^{2}+\tilde{k}_{-}^{2}+m^{2}}.  \label{E-}
\end{equation}%
The normalization constant is determined by the relation
\begin{equation}
|C_{\sigma }^{(-)}|^{2}=\frac{q\lambda (E_{-}-m)}{8\pi LE_{-}}\ .  \label{C-}
\end{equation}%
The wave-functions obtained in this section can be used for the
investigation of vacuum fermionic current densities induced by the presence
of the magnetic flux and also by the compactification along the string's
axis.

In the discussion above we have imposed the regularity condition on the
fermionc wave-functions at the cone apex. As it is well known, the theory of
von Neumann deficiency indices leads to a one-parameter family of allowed
boundary conditions in the background of an Aharonov-Bohm gauge field \cite%
{Sous89}. In addition to the regular modes, these boundary conditions, in
general, allow normalizable irregular modes. The VEV of the fermionic
current density for general boundary conditions on the cone apex is
evaluated in a way similar to that described below. The contribution of the
regular modes is the same for all boundary conditions and the results differ
by the parts related to the irregular modes. A special case of boundary
conditions has been discussed in \cite{Bene00}, where the
Atiyah-Patodi-Singer type nonlocal boundary condition is imposed at a finite
radius, which is then taken to zero. Similar approach, with the MIT bag
boundary condition, has been used in Refs. \cite{Saha10,Bell11} for a
two-dimensional conical space with a circular boundary. Note that in recent
investigation of the induced fermionic current for a massless Dirac field in
(2+1) dimensions, carried out in \cite{Jack09}, the authors impose the
regularity condition. It was shown that the corresponding result coincides
with the result for a finite radius solenoid, assuming that an electron
cannot penetrate the region of nonzero magnetic field.

\section{Fermionic current}

\label{sec3}

The VEV of the fermionic current density, $j^{\mu }=e\bar{\psi}\gamma ^{\mu
}\psi $, can be evaluated by using the mode sum formula,
\begin{equation}
\langle j^{\mu }(x)\rangle =e\sum_{\sigma }\bar{\psi}_{\sigma
}^{(-)}(x)\gamma ^{\mu }\psi _{\sigma }^{(-)}(x)\ ,  \label{current}
\end{equation}%
where we are using the compact notation defined below,
\begin{equation}
\sum_{\sigma }=\int_{0}^{\infty }d\lambda \ \sum_{l=-\infty }^{+\infty
}\sum_{s=\pm 1}\sum_{j=\pm 1/2,\cdots }\ .  \label{Sumsig}
\end{equation}%
This VEV\ is a periodic function of the fluxes $\Phi _{2}$ and $\Phi _{3}$
with the period equal to the flux quantum. In particular, if we write the
parameter $\alpha $ in (\ref{alfa}) in the form%
\begin{equation}
\alpha =n_{0}+\alpha _{0},\;|\alpha _{0}|<1/2,  \label{alfa0}
\end{equation}%
where $n_{0}$ is an integer number, the VEV of the current density will
depend on $\alpha _{0}$ only. Note that, for the boundary condition at the
cone apex used in \cite{Saha10}, there are no square integrable irregular
modes for $|\alpha _{0}|\leqslant (1-1/q)/2$.

\subsection{Charge density and radial current}

Let us start the calculation of charge density,
\begin{equation}
\rho (x)=\langle j^{0}(x)\rangle =e\sum_{\sigma }\psi _{\sigma }^{(-)\dagger
}\psi _{\sigma }^{(-)}\ .  \label{density}
\end{equation}%
Substituting (\ref{psi-}) and (\ref{C-}) into (\ref{density}), we obtain%
\begin{equation}
\rho (x)=\frac{eq}{4\pi L}\sum_{\sigma }\ \lambda \left[ J_{\beta
_{j}}^{2}(\lambda r)+J_{\beta _{j}+\epsilon _{j}}^{2}(\lambda r)\right] \ .
\end{equation}%
Of course, this expression is divergent. In order to regularize it we
introduce a cutoff function $e^{-\eta (\lambda ^{2}+k_{l}^{(-)2})}$, with
the cutoff parameter $\eta >0$. At the end of the calculation we take the
limit $\eta \rightarrow 0$. With the cutoff function, the integral can be
evaluated using the result from \cite{Grad}. So, the regularized
contribution due to the integral over $\lambda $ gives us:
\begin{equation}
\int_{0}^{\infty }d\lambda \ \lambda \ e^{-\eta \lambda ^{2}}\left[ J_{\beta
_{j}}^{2}(\lambda r)+J_{\beta _{j}+\epsilon _{j}}^{2}(\lambda r)\right] =%
\frac{1}{2\eta }e^{-r^{2}/(2\eta )}\left[ I_{\beta _{j}}(r^{2}/(2\eta
))+I_{\beta _{j}+\epsilon _{j}}(r^{2}/(2\eta ))\right] \ ,  \label{Int-reg}
\end{equation}%
with $I_{\nu }(z)$ being the modified Bessel function. As a result, the
regularized charge density reads:
\begin{equation}
\rho _{\mathrm{reg}}(x,\eta )=\frac{eqe^{-r^{2}/(2\eta )}}{4\pi \eta L}%
\sum_{l=-\infty }^{+\infty }e^{-\eta k_{l}^{(-)2}}\left[ \mathcal{I}%
(q,-\alpha _{0},r^{2}/(2\eta ))+\mathcal{I}(q,\alpha _{0},r^{2}/(2\eta ))%
\right] \ .
\end{equation}%
Here the expression for the regularized charge density is given in terms of
the series%
\begin{equation}
\mathcal{I}(q,\alpha _{0},z)=\sum_{j}I_{\beta _{j}}(z)=\sum_{n=0}^{\infty }%
\left[ I_{q(n+\alpha _{0}+1/2)-1/2}(z)+I_{q(n-\alpha _{0}+1/2)+1/2}(z)\right]
,  \label{seriesI0}
\end{equation}%
and
\begin{equation}
\sum_{j}I_{\beta _{j}+\epsilon _{j}}(z)=\mathcal{I}(q,-\alpha _{0},z)\ .
\label{seriesI2}
\end{equation}%
Here and below%
\begin{equation}
\sum_{j}=\sum_{j=\pm 1/2,\cdots }.
\end{equation}

An equivalent representation for the charge density can be obtained by using
the integral representation below derived in \cite{Saha10}:
\begin{eqnarray}
&&\mathcal{I}(q,\alpha _{0},z)=\frac{e^{z}}{q}-\frac{1}{\pi }%
\int_{0}^{\infty }dy\frac{e^{-z\cosh y}f(q,\alpha _{0},y)}{\cosh (qy)-\cos
(q\pi )}  \notag \\
&&\qquad +\frac{2}{q}\sum_{k=1}^{p}(-1)^{k}\cos [2\pi k(\alpha
_{0}-1/2q)]e^{z\cos (2\pi k/q)},  \label{seriesI3}
\end{eqnarray}%
with $2p<q<2p+2$ and with the notation%
\begin{eqnarray}
f(q,\alpha _{0},y) &=&\cos \left[ q\pi \left( 1/2-\alpha _{0}\right) \right]
\cosh \left[ \left( q\alpha _{0}+q/2-1/2\right) y\right]  \notag \\
&&-\cos \left[ q\pi \left( 1/2+\alpha _{0}\right) \right] \cosh \left[
\left( q\alpha _{0}-q/2-1/2\right) y\right] \ .  \label{fqualf}
\end{eqnarray}%
In the case $q=2p$, the term
\begin{equation}
-(-1)^{q/2}\frac{e^{-z}}{q}\sin (\pi q\alpha _{0})\ ,  \label{replaced}
\end{equation}%
should be added to the right-hand side of Eq. (\ref{seriesI3}). For $%
1\leqslant q<2$, the last term on the right-hand side of Eq. (\ref{seriesI3}%
) is absent.

Note that for integer values of $q$ and for%
\begin{equation}
\alpha _{0}=\frac{1}{2}-\frac{n+1/2}{q}\ ,  \label{gammmaSp}
\end{equation}%
with an integer $n$, one has $f(q,\alpha _{0},y)=0$. From the condition $%
|\alpha _{0}|<1/2$ we find%
\begin{equation}
0\leqslant n<q-1/2\ .  \label{nBound}
\end{equation}%
In this case we can see that the functions $\mathcal{I}(q,\pm \alpha _{0},z)$
are presented in equivalent forms:%
\begin{eqnarray}
\mathcal{I}(q,\alpha _{0},z) &=&\frac{1}{q}\sum_{k=0}^{q-1}\cos (2\pi
k(n+1)/q)e^{z\cos (2\pi k/q)}\ ,  \notag \\
\mathcal{I}(q,-\alpha _{0},z) &=&\frac{1}{q}\sum_{k=0}^{q-1}\cos (2\pi
kn/q)e^{z\cos (2\pi k/q)}.  \label{ISp}
\end{eqnarray}

For a general $q$, by making use of the formula (\ref{seriesI3}), the
regularized charge density is presented in the form
\begin{eqnarray}
\rho _{\mathrm{reg}}(x,\eta ) &=&\frac{ee^{-z}}{2\pi \eta L}\sum_{l=-\infty
}^{+\infty }e^{-\eta k_{l}^{(-)2}}\left[ e^{z}+\frac{q}{\pi }%
\int_{0}^{\infty }dy\frac{\sinh \left( y/2\right) e^{-z\cosh y}h(q,\alpha
_{0},y)}{\cosh (qy)-\cos (q\pi )}\right.   \notag \\
&&\left. +2\sum_{k=1}^{p}(-1)^{k}c_{k}\cos \left( 2\pi k\alpha _{0}\right)
e^{z\cos (2\pi k/q)}\right] \ ,  \label{roReg}
\end{eqnarray}%
with $z=r^{2}/(2\eta )$. In this representation we have introduced the
notations%
\begin{equation}
c_{k}=\cos \left( \pi k/q\right) ,  \label{ck}
\end{equation}%
and%
\begin{eqnarray}
h(q,\alpha _{0},y) &=&\cos \left[ q\pi \left( 1/2+\alpha _{0}\right) \right]
\sinh \left[ q\left( 1/2-\alpha _{0}\right) y\right]   \notag \\
&&+\cos \left[ q\pi \left( 1/2-\alpha _{0}\right) \right] \sinh \left[
q\left( 1/2+\alpha _{0}\right) y\right] \ .  \label{g0}
\end{eqnarray}%
The first term in the square brackets of (\ref{roReg}) corresponds to the
charge density for $\alpha _{0}=0$ and $q=1$. The renormalized value for
this part vanishes (see Ref. \cite{Bell10} for a general case of spatial
topology $R^{p}\times (S^{1})^{q}$ and Ref. \cite{Bell13} for the
corresponding current densities in de Sitter spacetime). The other
contributions contain $e^{-r^{2}\cosh ^{2}(y/2)/\eta }$ and $e^{-r^{2}\sin
^{2}(\pi k/q)/\eta }$, inside the integral and summation respectively; hence
in the limit $\eta \rightarrow 0$ these terms vanish for $r>0$. So, we
conclude that the renormalized value for the charge density is zero, i.e,
there is no induced charge density.

As to the VEV of the radial current, it is given by the expression
\begin{equation}
\langle j^{r}(x)\rangle =e\sum_{\sigma }\psi _{\sigma }^{(-)\dagger }\gamma
^{0}\gamma ^{r}\psi _{\sigma }^{(-)}\ .  \label{Radialcurr}
\end{equation}%
Substituting (\ref{psi-}) and the Dirac matrices given in (\ref{gamcurved})
and (\ref{betl}) in the right-hand side of (\ref{Radialcurr}), we can easily
see that there appears a cancellation between all terms. Consequently there
is also no induced radial current density.

\subsection{Azimuthal current}

The VEV of the azimuthal current is given by
\begin{equation}
\langle j^{\phi }(x)\rangle =e\sum_{\sigma }\psi _{\sigma }^{(-)\dagger
}\gamma ^{0}\gamma ^{\phi }\psi _{\sigma }^{(-)}\ .  \label{Azimucurr}
\end{equation}%
Substituting the expression for the negative-energy solution for the
fermionic field (\ref{psi-}) and the corresponding expressions for the Dirac
matrices in this background, given in (\ref{gamcurved}) and (\ref{betl}),
into the expression inside the summations of (\ref{Azimucurr}), after the
redefinition $l\rightarrow -l$, we obtain:
\begin{equation}
\langle j^{\phi }\rangle =-\frac{eq}{2\pi Lr}\sum_{\sigma }\frac{\lambda
^{2}\epsilon _{j}J_{\beta _{j}}(\lambda r)J_{\beta _{j}+\epsilon
_{j}}(\lambda r)}{\sqrt{[2\pi (l-\tilde{\beta})/L]^{2}+\lambda ^{2}+m^{2}}}\
.  \label{Azimucurr1}
\end{equation}%
We assume the presence of a cutoff function without writing it explicitly.
The specific form of this function is not needed in the further discussion.

The summation over the quantum number $s$ in (\ref{Azimucurr1}) provides a
factor $2$. In order to develop the summation over $l$, we shall apply the
Abel-Plana summation formula in the form \cite{Bell10} (for generalizations
of the Abel-Plana formula see \cite{SahaBook})
\begin{eqnarray}
&&\sum_{l=-\infty }^{\infty }g(l+\tilde{\beta})f(|l+\tilde{\beta}%
|)=\int_{0}^{\infty }du\,\left[ g(u)+g(-u)\right] f(u)  \notag \\
&&\qquad +i\int_{0}^{\infty }du\left[ f(iu)-f(-iu)\right] \sum_{\lambda =\pm
1}\frac{g(i\lambda u)}{e^{2\pi (u+i\lambda \tilde{\beta})}-1}\ ,
\label{sumform}
\end{eqnarray}%
taking $g(u)=1$ and
\begin{equation}
f(u)=\frac{1}{\sqrt{(2\pi u/L)^{2}+\lambda ^{2}+m^{2}}}\ .  \label{fu}
\end{equation}%
As a result, the induced azimuthal current is decomposed as,
\begin{equation}
\langle j^{\phi }\rangle =\langle j^{\phi }\rangle _{s}+\langle j^{\phi
}\rangle _{c}\ ,
\end{equation}%
where the term $\langle j^{\phi }\rangle _{s}$ is due to the contribution of
the first integral in the right-hand side of (\ref{sumform}) and corresponds
to the axial current density in the geometry of a cosmic string without
compactification. The part $\langle j^{\phi }\rangle _{c}$ is induced by the
compactification of the string along its axis. As we shall see the latter
vanishes in the limit $L\rightarrow \infty $.

The calculation of the induced azimuthal current in the geometry of a
straight cosmic string has been developed before by many authors considering
massless field. For massive field the expression is provided in a closed
form for the special case where $q$ is an integer and $\alpha _{0}$ given by
(\ref{gammmaSp}) with $n=0$ \cite{Mello10}; however, to our knowledge, a
closed expression for the induced azimuthal current considering general
values of the parameters is missed. So, in order to fill this blank, we
decided to include this calculation in the present paper. Combining (\ref%
{Azimucurr1}) and (\ref{sumform}), we get the representation
\begin{equation}
\langle j^{\phi }\rangle _{s}=-\frac{eq}{\pi ^{2}r}\int_{0}^{\infty
}d\lambda \lambda ^{2}\int_{0}^{\infty }dk\sum_{j}\frac{\epsilon
_{j}J_{\beta _{j}}(\lambda r)J_{\beta _{j}+\epsilon _{j}}(\lambda r)}{\sqrt{%
m^{2}+k^{2}+\lambda ^{2}}}\ .  \label{j-cs}
\end{equation}%
In order to provide a more workable expression, we use the identity
\begin{equation}
\frac{1}{\sqrt{m^{2}+k^{2}+\lambda ^{2}}}=\frac{2}{\sqrt{\pi }}%
\int_{0}^{\infty }dt\ e^{-(m^{2}+k^{2}+\lambda ^{2})t^{2}}\ .  \label{ident1}
\end{equation}%
Substituting this identity into (\ref{j-cs}), the next step is to develop
the integral over the variable $\lambda $. With the help of \cite{Grad}, we
can write,
\begin{equation}
\int_{0}^{\infty }d\lambda \lambda ^{2}e^{-\lambda ^{2}t^{2}}J_{\beta
_{j}}(\lambda r)J_{\beta _{j}+\epsilon _{j}}(\lambda r)=\frac{%
e^{-r^{2}/(2t^{2})}}{4t^{4}}r\epsilon _{j}\left[ I_{\beta
_{j}}(r^{2}/(2t^{2}))-I_{\beta _{j}+\epsilon _{j}}(r^{2}/(2t^{2}))\right] \ .
\label{Int1}
\end{equation}%
Introducing a new variable $y=r^{2}/(2t^{2})$, we obtain
\begin{equation}
\langle j^{\phi }\rangle _{s}=\frac{eq}{2\pi ^{2}r^{4}}\int_{0}^{\infty }\
dy\ y\ e^{-y-m^{2}r^{2}/(2y)}\ [\mathcal{I}(q,-\alpha _{0},y)-\mathcal{I}%
(q,\alpha _{0},y)]\ ,  \label{j-cs1}
\end{equation}%
where $\mathcal{I}(q,\alpha _{0},y)$ is defined in (\ref{seriesI0}). From
the above expression, we see that the induced azimuthal current is an odd
function of $\alpha _{0}$.

By using the formula (\ref{seriesI3}), after the integration over $y$, the
expression (\ref{j-cs1}) is presented in the form%
\begin{eqnarray}
\langle j^{\phi }\rangle _{s} &=&-\frac{em^{2}}{\pi ^{2}r^{2}}\ \left[
\sum_{k=1}^{p}\frac{(-1)^{k}}{s_{k}}\sin (2\pi k\alpha
_{0})K_{2}(2mrs_{k})\right.   \notag \\
&&\left. +\frac{q}{\pi }\int_{0}^{\infty }dy\frac{g(q,\alpha
_{0},2y)K_{2}(2mr\cosh y)}{[\cosh (2qy)-\cos (q\pi )]\cosh y}\right] \ ,
\label{jazimu}
\end{eqnarray}%
where $K_{\nu }(x)$ is the Macdonald function. In (\ref{jazimu}), we have
introduced the notations%
\begin{equation}
s_{k}=\sin (\pi k/q),  \label{sk}
\end{equation}%
and%
\begin{eqnarray}
g(q,\alpha _{0},y) &=&\cos \left[ q\pi \left( 1/2+\alpha _{0}\right) \right]
\cosh \left[ q\left( 1/2-\alpha _{0}\right) y\right]   \notag \\
&&-\cos \left[ q\pi \left( 1/2-\alpha _{0}\right) \right] \cosh \left[
q\left( 1/2+\alpha _{0}\right) y\right] .  \label{gxy}
\end{eqnarray}%
The azimuthal current density $\langle j^{\phi }\rangle _{s}$ vanishes in
the absence of the magnetic flux along the string ($\alpha _{0}=0$). In the
special case $q=1$, the first term in the square brackets of (\ref{jazimu})
is absent and from this formula we obtain the current density in the
Minkowski bulk induced by the magnetic flux.

At large distances from the string, $mr\gg 1$, and for $q\geqslant 2$ the
dominant contribution to (\ref{jazimu}) comes from the term with $k=1$. To
the leading order one gets%
\begin{equation}
\langle j^{\phi }\rangle _{s}\approx \frac{em^{4}\sin (2\pi \alpha _{0})}{%
2(\pi s_{1})^{3/2}(mr)^{5/2}}\ e^{-2mrs_{1}}.  \label{jslarge}
\end{equation}%
For $q<2$ and $mr\gg 1$, the azimuthal current density is suppressed by the
factor $e^{-2mr}$. Near the string, $mr\ll 1$, the leading term in $\langle
j^{\phi }\rangle _{s}$ behaves as $1/r^{4}$. This term does not depend on
the mass and coincides with $\langle j^{\phi }\rangle _{s}$ for a massless
field. The latter is easily obtained from (\ref{jazimu}) by taking into
account that $K_{\nu }(x)\sim 2^{\nu -1}\Gamma (\nu )x^{-\nu }$ for $%
x\rightarrow 0$.

For the special case (\ref{gammmaSp}) the integral term in (\ref{jazimu})
vanishes and one finds:
\begin{equation}
\langle j^{\phi }\rangle _{s}=\frac{em^{2}}{2\pi ^{2}r^{2}}\
\sum_{k=1}^{q-1}\sin \left( \pi k\frac{2n+1}{q}\right) \frac{K_{2}(2mrs_{k})%
}{s_{k}}\ .  \label{jazimu1}
\end{equation}%
For a massless field and for $n=0$, it can be seen that (\ref{jazimu1})
becomes,
\begin{equation}
\langle j^{\phi }\rangle _{s}=\frac{e(q^{2}-1)}{12\pi ^{2}r^{4}}\ .
\end{equation}

\begin{figure}[tbph]
\begin{center}
\epsfig{figure=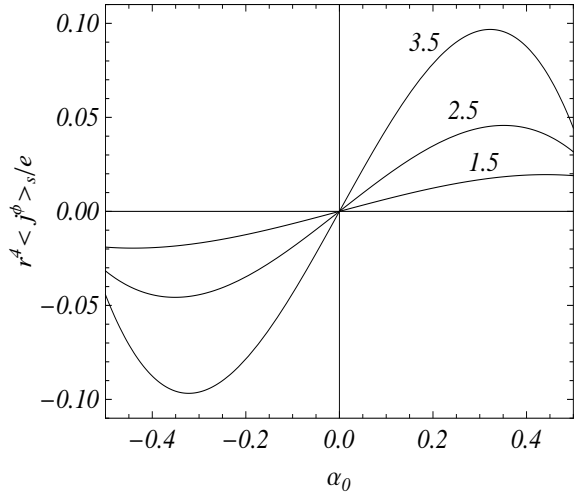,width=7.5cm,height=6.5cm}
\end{center}
\caption{Azimuthal current in the geometry of a straight cosmic string, $%
r^{4}\langle j^{\protect\phi }\rangle _{s}/e$, as a function of the
parameter $\protect\alpha _{0}$ in the case of a massless fermionic field
for different values of $q$ (numbers near the curves). }
\label{fig1}
\end{figure}

In fig. \ref{fig1} we display the dependence of $r^{4}\langle j^{\phi
}\rangle _{s}/e$ on $\alpha _{0}$ in the case of a massless fermionic field
for separate values of the parameter $q$ (numbers near the curves). As it is
seen, the current density increases with increasing $q$.

Now let us develop the calculation of the contribution to the azimuthal
current induced by the compactification of the string along its axis. This
part comes from the second integral in the right-hand side of the summation
formula (\ref{sumform}) and is presented in the form:
\begin{eqnarray}
\langle j^{\phi }\rangle _{c} &=&-\frac{eq}{\pi ^{2}r}\sum_{j}\epsilon
_{j}\int_{0}^{\infty }d\lambda \lambda ^{2}J_{\beta _{j}}(\lambda r)J_{\beta
_{j}+\epsilon _{j}}(\lambda r)\int_{\sqrt{\lambda ^{2}+m^{2}}}^{\infty }dk
\notag \\
&&\times \frac{1}{\sqrt{k^{2}-\lambda ^{2}-m^{2}}}\left( \frac{1}{e^{Lk+2\pi
i\tilde{\beta}}-1}+\frac{1}{e^{Lk-2\pi i\tilde{\beta}}-1}\right) \ .
\label{jc-1}
\end{eqnarray}%
To continue the calculation, we shall use the series expansion, $\left(
e^{u}-1\right) ^{-1}=\sum_{l=1}^{\infty }e^{-lu}$ in the above expression.
Taking this expansion, the integral over $k$ can be developed with the help
of \cite{Grad}. After some minor steps, we arrive at,
\begin{equation}
\langle j^{\phi }\rangle _{c}=-\frac{2eq}{\pi ^{2}r}\sum_{l=1}^{\infty }\cos
(2\pi l\tilde{\beta})\sum_{j}\epsilon _{j}\int_{0}^{\infty }d\lambda \lambda
^{2}J_{\beta _{j}}(\lambda r)J_{\beta _{j}+\epsilon _{j}}(\lambda r)K_{0}(lL%
\sqrt{m^{2}+\lambda ^{2}})\ .
\end{equation}
At this point we shall use the integral representation below for the
Macdonald function \cite{Grad}:
\begin{equation}
K_{\nu }(x)=\frac{1}{2}\left( \frac{x}{2}\right) ^{\nu }\int_{0}^{\infty }dt%
\frac{e^{-t-x^{2}/(4t)}}{t^{\nu +1}}\ .  \label{Kintergral}
\end{equation}%
So, we obtain the representation
\begin{eqnarray}
\langle j^{\phi }\rangle _{c} &=&-\frac{eq}{\pi ^{2}r}\sum_{l=1}^{\infty
}\cos (2\pi l\tilde{\beta})\sum_{j}\epsilon _{j}\int_{0}^{\infty }dt\ \frac{%
e^{-t-l^{2}L^{2}m^{2}/(4t)}}{t}  \notag \\
&&\times \int_{0}^{\infty }d\lambda \lambda ^{2}J_{\beta _{j}}(\lambda
r)J_{\beta _{j}+\epsilon _{j}}(\lambda r)e^{-l^{2}L^{2}\lambda ^{2}/4t}\ .
\label{jc-3}
\end{eqnarray}
The integral over $\lambda $ can be developed by using a similar integral as
is written in (\ref{Int1}). Defining a new variable $y=2r^{2}t/(l^{2}L^{2})$%
, and after some simplifications, we arrive at,
\begin{eqnarray}
\langle j^{\phi }\rangle _{c} &=&\frac{eq}{\pi ^{2}r^{4}}\sum_{l=1}^{\infty
}\cos (2\pi l\tilde{\beta})\int_{0}^{\infty }dy\
ye^{-y[1+l^{2}L^{2}/(2r^{2})]-m^{2}r^{2}/(2y)}  \notag \\
&&\times \lbrack \mathcal{I}(q,-\alpha _{0},y)-\mathcal{I}(q,\alpha
_{0},y)]\ .  \label{jc-4}
\end{eqnarray}%
We can see that this contribution remains the same for the replacement of $%
\tilde{\beta}$ by $1-\tilde{\beta}$.

For the further evaluation of the topological part we use the representation
(\ref{seriesI3}) for the functions $\mathcal{I}(q,\pm \alpha _{0},y)$. After
the integration over $y$ we come to the expression%
\begin{eqnarray}
\langle j^{\phi }\rangle _{c} &=&-\frac{8em^{2}}{\pi ^{2}L^{2}}%
\sum_{l=1}^{\infty }\cos (2\pi l\tilde{\beta})\left[
\sum_{k=1}^{p}(-1)^{k}s_{k}\sin (2\pi k\alpha _{0})\frac{K_{2}(mL\sqrt{\rho
_{k}^{2}+l^{2}})}{\rho _{k}^{2}+l^{2}}\right.  \notag \\
&&\left. +\frac{q}{\pi }\int_{0}^{\infty }dy\frac{\cosh (y)g(q,\alpha
_{0},2y)}{\cosh (2qy)-\cos (q\pi )}\frac{K_{2}(mL\sqrt{\eta ^{2}(y)+l^{2}})}{%
\eta ^{2}(y)+l^{2}}\right] \ ,  \label{jc-4b}
\end{eqnarray}%
where we have defined%
\begin{equation}
\rho _{k}=\frac{2rs_{k}}{L}\ ,\;\eta (y)=\frac{2r\cosh y}{L}.  \label{rho-k}
\end{equation}%
As we see, the part in the current density induced by the compactification
of the string axis is an odd function of the magnetic flux along the string
and it is an even function of the parameter $\tilde{\beta}$. In particular,
in the case of an untwisted fermionic field, $\langle j^{\phi }\rangle _{c}$
is an even function of the magnetic flux enclosed by the string axis. In the
absence of the magnetic flux along the string axis one has $\langle j^{\phi
}\rangle _{c}=0$.

The topological part of the current density is finite on the string and from
(\ref{jc-4b}) we get
\begin{eqnarray}
\langle j^{\phi }\rangle _{c}|_{r=0} &=&-\frac{8em^{2}}{\pi ^{2}L^{2}}%
\sum_{l=1}^{\infty }\cos (2\pi l\tilde{\beta})\frac{K_{2}(mLl)}{l^{2}}
\notag \\
&&\times \left[ \sum_{k=1}^{p}(-1)^{k}s_{k}\sin (2\pi k\alpha _{0})+\frac{q}{%
\pi }\int_{0}^{\infty }dy\frac{\cosh (y)g(q,\alpha _{0},2y)}{\cosh
(2qy)-\cos (q\pi )}\right] \ .  \label{jcNearStr}
\end{eqnarray}%
Hence, near the string the total current is dominated by the part $\langle
j^{\phi }\rangle _{s}$. For large values of the length of the compact
dimension, $mL\gg 1$, assuming that $mr$ is fixed, the main contribution
comes from the $l=1$ term and to the leading order we find%
\begin{eqnarray}
\langle j^{\phi }\rangle _{c} &\approx &-\frac{2^{5/2}em^{3/2}\cos (2\pi
\tilde{\beta})e^{-mL}}{\pi ^{3/2}L^{5/2}}\left[ \sum_{k=1}^{p}(-1)^{k}s_{k}%
\sin (2\pi k\alpha _{0})\right.  \notag \\
&&\left. +\frac{q}{\pi }\int_{0}^{\infty }dy\frac{\cosh (y)g(q,\alpha
_{0},2y)}{\cosh (2qy)-\cos (q\pi )}\right] \ .  \label{jcLlarge}
\end{eqnarray}%
In this limit the dominant contribution to the total current density comes
from the term $\langle j^{\phi }\rangle _{s}$.

In the special case (\ref{gammmaSp}) the formula (\ref{jc-4b}) is simplified
to%
\begin{equation}
\langle j^{\phi }\rangle _{c}=\frac{4em^{2}}{\pi ^{2}L^{2}}%
\sum_{l=1}^{\infty }\cos (2\pi l\tilde{\beta })\sum_{k=1}^{q-1}s_{k}\sin
\left( \pi k\frac{2n+1}{q}\right) \frac{K_{2}(mL\sqrt{\rho _{k}^{2}+l^{2}})}{%
\rho _{k}^{2}+l^{2}}\ ,  \label{jcSp}
\end{equation}%
where $n$ is restricted by the condition (\ref{nBound}).

For a massless field, combining (\ref{jc-4b}) with (\ref{jazimu}), the
expression for the total azimuthal current density takes the form
\begin{eqnarray}
\langle j^{\phi }\rangle &=&-\frac{16e}{\pi ^{2}L^{4}}\left[
\sum_{k=1}^{p}(-1)^{k}s_{k}\sin (2\pi k\alpha _{0})C(\tilde{\beta},\rho
_{k})\right.  \notag \\
&&\left. +\frac{q}{\pi }\int_{0}^{\infty }dy\frac{g(q,\alpha _{0},2y)\cosh y%
}{\cosh (2qy)-\cos (q\pi )}C(\tilde{\beta},\eta (y))\right] \ \ ,
\label{jcm0}
\end{eqnarray}%
where we have defined%
\begin{equation}
C(\tilde{\beta},x)=\sideset{}{'}{\sum}_{l=0}^{\infty }\frac{\cos (2\pi l%
\tilde{\beta})}{\left( l^{2}+x^{2}\right) ^{2}},  \label{Cbet}
\end{equation}%
and the prime on the sum means that the term $l=0$ should be taken with the
coefficient 1/2. In (\ref{jcm0}), the part with $l=0$ in (\ref{Cbet})
coincides with $\langle j^{\phi }\rangle _{s}$ and the remaining part
corresponds to $\langle j^{\phi }\rangle _{c}$. Note that for the series (%
\ref{Cbet}) one has \cite{Prud86}%
\begin{eqnarray}
C(\tilde{\beta},x) &=&\frac{\pi ^{2}\cosh (2\pi \tilde{\beta}x)}{4x^{2}\sinh
^{2}(\pi x)}  \notag \\
&&+\pi \frac{\cosh [\pi (1-2\tilde{\beta})x]+2\pi \tilde{\beta}x\sinh [\pi
(1-2\tilde{\beta})x]}{4x^{3}\sinh (\pi x)},  \label{Cbet1}
\end{eqnarray}%
where $0\leqslant \tilde{\beta}\leqslant 1$.

\begin{figure}[tbph]
\begin{center}
\epsfig{figure=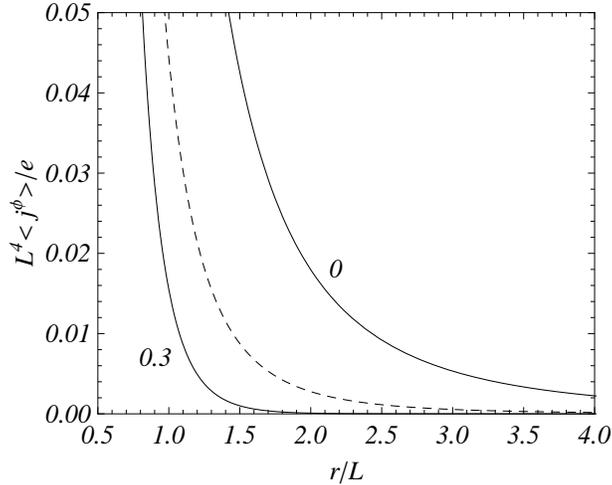,width=8.cm,height=6.5cm}
\end{center}
\caption{The total azimuthal current density, $L^{4}\langle j^{\protect\phi %
}\rangle /e$, for a massless fermionic field as a function of $r/L$ for $%
q=2.5$ and $\protect\alpha _{0}=0.4$. The numbers near the curves are the
values of the parameter $\tilde{\protect\beta}$ and the dashed curve
corresponds to the current density for the geometry of a straight cosmic
string, $L^{4}\langle j^{\protect\phi }\rangle _{s}/e$.}
\label{fig2}
\end{figure}

For $r\ll L$ and for a massless field, for the topological part to the
leading order one has%
\begin{eqnarray}
\langle j^{\phi }\rangle _{c} &\approx &\frac{16\pi ^{2}e}{3L^{4}}\left[
\tilde{\beta}^{2}(1-\tilde{\beta})^{2}-\frac{1}{30}\right] \left[
\sum_{k=1}^{p}(-1)^{k}s_{k}\sin (2\pi k\alpha _{0})\right.   \notag \\
&&\left. +\frac{q}{\pi }\int_{0}^{\infty }dy\frac{\cosh (y)g(q,\alpha
_{0},2y)}{\cosh (2qy)-\cos (q\pi )}\right] \ \ ,  \label{jcm0Smr}
\end{eqnarray}%
where we have assumed that $0\leqslant \tilde{\beta}\leqslant 1$. Note that
in this limit $\langle j^{\phi }\rangle _{c}/\langle j^{\phi }\rangle
_{s}\sim (r/L)^{4}$, and the total current density is dominated by the part $%
\langle j^{\phi }\rangle _{s}$. In the opposite limit, $r\gg L$, we use the
asymptotic expression for the function (\ref{Cbet1}) for $x\gg 1$ and for $0<%
\tilde{\beta}<1$: $C(\tilde{\beta},x)\approx \pi ^{2}\sigma e^{-2\pi \sigma
x}/(2x^{2})$, where $\sigma =\min (\tilde{\beta},1-\tilde{\beta})$. For $%
\tilde{\beta}=0$ one has the asymptotic $C(0,x)\approx \pi /(4x^{3})$. From
these expressions we conclude that at large distances from the string, $r\gg
L$, the azimuthal current density is exponentially suppressed by the factor $%
\exp [-4\pi \sigma r\sin (\pi /q)/L]$ for $q\geqslant 2$ and for $0<\tilde{%
\beta}<1$. For $q<2$ the suppression is by the factor $\exp [-4\pi \sigma
r/L]$. For $\tilde{\beta}=0$ the current density decays as power-law: $%
\langle j^{\phi }\rangle \sim (L/r)^{3}$. Note that in the latter case the
total current density is dominated by the topological part: $\langle j^{\phi
}\rangle _{c}/\langle j^{\phi }\rangle _{s}\sim r/L$. Hence, at distances
larger than the length of the compactification the behavior of the azimuthal
current density depends crucially on whether $\tilde{\beta}=0$ or not. For $%
\tilde{\beta}\neq 0$ the compactification of the string along its axis leads
to the suppression of the current density, whereas for $\tilde{\beta}=0$ the
current density is increased by the compactification. This feature is
illustrated in figure \ref{fig2}, where we display the total azimuthal
current density for a massless fermionic field as a function of $r/L$. The
graphs are plotted for $q=2.5$, $\alpha _{0}=0.4$ and the numbers near the
curves correspond to the values of the parameter $\tilde{\beta}$. The dashed
curve presents the current density for the geometry of a straight cosmic
string ($L^{4}\langle j^{\phi }\rangle _{s}/e$).

\begin{figure}[tbph]
\begin{center}
\epsfig{figure=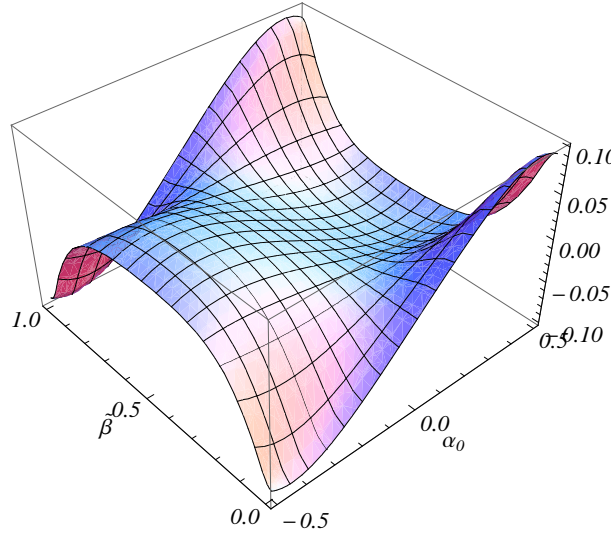,width=8.cm,height=7.cm}
\end{center}
\caption{The topological part of the azimuthal current density, $%
L^{4}\langle j^{\protect\phi }\rangle _{c}/e$, as a function of $\protect%
\alpha _{0}$ and $\tilde{\protect\beta}$ for $q=2.5$ and $r/L=1$.}
\label{fig3}
\end{figure}

In fig. \ref{fig3} we plot the topological part of the azimuthal current
density, $L^{4}\langle j^{\phi }\rangle _{c}/e$, as a function of $\alpha
_{0}$ and $\tilde{\beta}$ in the geometry of a cosmic string with the
parameter $q=2.5$ and for $r/L=1$.

\section{Axial current}

\label{sec4}

In this section we shall present the evaluation of the induced fermionic
current along the string's axis. As we shall see, this new current is a
consequence of the quasiperiodicity condition imposed on the fermionic field
and also of the magnetic flux enclosed by the compact dimension.

The VEV of the axial current is given by the expression
\begin{equation}
\langle j^{z}\rangle =e\sum_{\sigma }\psi _{\sigma }^{(-)\dagger }\gamma
^{0}\gamma ^{z}\psi _{\sigma }^{(-)}\ .  \label{Axial}
\end{equation}%
Once more, substituting the expression for the negative-energy mode
functions (\ref{psi-}) and the corresponding expressions for the Dirac
matrices in this background, given in (\ref{gamcurved}), into the expression
in the right-hand side of (\ref{Axial}), we obtain:
\begin{equation}
\langle j^{z}\rangle =-\frac{eq}{4\pi L}\sum_{\sigma }\lambda \frac{%
k_{l}^{(-)}}{E_{-}}[J_{\beta _{j}}^{2}(\lambda r)+J_{\beta _{j}+\epsilon
_{j}}^{2}(\lambda r)]\ ,  \label{Axial-1}
\end{equation}%
with $\sum_{\sigma }$ defined as (\ref{Sumsig}). The summation over $s$
provides the factor $2$. As to the summation over the quantum number $l$, we
redefine $l\rightarrow -l$ (this is equivalent to the replacements $%
k_{l}^{(-)}\rightarrow -k_{l}^{(+)}$ and $E_{-}\rightarrow E_{+}$ in (\ref%
{Axial-1})) and then use the formula (\ref{sumform}), taking $g(u)=2\pi u/L$
and the expression (\ref{fu}) for the function $f(u)$. For this case, the
function $g(u)$ is an odd function and the contribution due to the first
term on the right-hand side of (\ref{sumform}) vanishes, remaining only the
contribution associated with the second one, which will provide the current
due to the compactification. As a consequence, the axial current density in
the geometry of an uncompactified cosmic string vanishes.

The current density induced by the compactification of the string along its
axis is written as,
\begin{eqnarray}
\langle j^{z}\rangle _{c} &=&\frac{iqe}{2\pi ^{2}}\sum_{j}\int_{0}^{\infty
}d\lambda \ \lambda \lbrack J_{\beta _{j}}^{2}(\lambda r)+J_{\beta
_{j}+\epsilon _{j}}^{2}(\lambda r)]\int_{\sqrt{\lambda ^{2}+m^{2}}}^{\infty
}dk  \notag \\
&&\times \frac{k}{\sqrt{k^{2}-\lambda ^{2}-m^{2}}}\left( \frac{1}{e^{Lk+2\pi
i\tilde{\beta}}-1}-\frac{1}{e^{Lk-2\pi i\tilde{\beta}}-1}\right) \ .
\label{Axial-2}
\end{eqnarray}%
As before, the next step is to use the expansion $\left( e^{u}-1\right)
^{-1}=\sum_{l=1}^{\infty }e^{-lu}$, in the terms inside the bracket, and
with the help of \cite{Grad}, the integral over $k$ can be evaluated with
the result:
\begin{eqnarray}
\langle j^{z}\rangle _{c} &=&\frac{qe}{\pi ^{2}}\sum_{l=1}^{\infty }\sin
(2\pi l\tilde{\beta})\sum_{j}\int_{0}^{\infty }d\lambda \ \lambda \sqrt{%
\lambda ^{2}+m^{2}}  \notag \\
&&\times \lbrack J_{\beta _{j}}^{2}(\lambda r)+J_{\beta _{j}+\epsilon
_{j}}^{2}(\lambda r)]K_{1}\left( lL\sqrt{\lambda ^{2}+m^{2}}\right) \ .
\end{eqnarray}%
As we see the induced axial current is an odd periodic function of $\tilde{%
\beta}$ with the period 1. Now using the integral representation for the
Macdonald function given in (\ref{Kintergral}), we arrive:
\begin{eqnarray}
\langle j^{z}\rangle _{c} &=&\frac{qeL}{4\pi ^{2}}\sum_{l=1}^{\infty }l\sin
(2\pi l\tilde{\beta})\int_{0}^{\infty }dt\ \frac{e^{-t-l^{2}L^{2}m^{2}/(4t)}%
}{t^{2}}\int_{0}^{\infty }d\lambda \,\lambda  \notag \\
&&\times (\lambda ^{2}+m^{2})e^{-l^{2}L^{2}\lambda
^{2}/(4t)}\sum_{j}[J_{\beta _{j}}^{2}(\lambda r)+J_{\beta _{j}+\epsilon
_{j}}^{2}(\lambda r)]\ .  \label{Axial-3}
\end{eqnarray}%
The integral over $\lambda $ presents two terms. The term proportional to $%
m^{2}$ is given in \cite{Grad} and the other term that contains the power $%
\lambda ^{3}$ can be evaluated by using a well-known trick. So, we have
\begin{eqnarray}
\int_{0}^{\infty }d\lambda \ \lambda (\lambda ^{2}+m^{2})\ e^{-a\lambda
^{2}}J_{\nu }^{2}(\lambda r) &=&\left( m^{2}-\partial _{a}\right)
\int_{0}^{\infty }d\lambda \ \lambda \ e^{-a\lambda ^{2}}J_{\nu
}^{2}(\lambda r)  \notag \\
&=&\frac{1}{2a^{2}}\left( m^{2}a+1+y\partial _{y}\right) \left[ e^{-y}I_{\nu
}\left( y\right) \right] \ ,
\end{eqnarray}%
with $y=r^{2}/(2a)$. Adapting the above expression to our calculation we
have,%
\begin{eqnarray}
\langle j^{z}\rangle _{c} &=&\frac{16qe}{8\pi ^{2}L^{3}}\sum_{l=1}^{\infty }%
\frac{\sin (2\pi l\tilde{\beta})}{l^{3}}\int_{0}^{\infty }dt\
e^{-t-l^{2}L^{2}m^{2}/(4t)}  \notag \\
&&\times \left( \frac{l^{2}L^{2}m^{2}}{4t}+1+y\partial _{y}\right) \left\{
e^{-y}\left[ \mathcal{I}(q,\alpha _{0},y)+\mathcal{I}(q,-\alpha _{0}\gamma
,y)\right] \right\} \ ,
\end{eqnarray}%
where $y=2tr^{2}/(lL)^{2}$. As it is seen, the axial current density is an
even function of $\alpha _{0}$.

Now we use the formula (\ref{seriesI3}) for the function $\mathcal{I}%
(q,\alpha _{0},y)$. After the integration over $t$ this gives%
\begin{eqnarray}
\langle j^{z}\rangle _{c} &=&\frac{4m^{2}e}{\pi ^{2}L}\sum_{l=1}^{\infty
}l\sin (2\pi l\tilde{\beta})\left[ \sideset{}{'}{\sum}%
_{k=0}^{p}(-1)^{k}c_{k}\cos \left( 2\pi k\alpha _{0}\right) \frac{K_{2}(mL%
\sqrt{l^{2}+\rho _{k}^{2}})}{l^{2}+\rho _{k}^{2}}\right.  \notag \\
&&\left. +\frac{q}{\pi }\int_{0}^{\infty }dy\frac{h(q,\alpha _{0},2y)\sinh y%
}{\cosh (2qy)-\cos (q\pi )}\frac{K_{2}(mL\sqrt{l^{2}+\eta ^{2}(y)})}{%
l^{2}+\eta ^{2}(y)}\right] \ ,  \label{jzc}
\end{eqnarray}%
where $c_{k}$ is defined in (\ref{ck}), the function $h(q,\alpha _{0},v)$ is
given in (\ref{g0}), and the prime on the sign of the sum means that the
term with $k=0$ should be taken with the coefficient 1/2. This term does not
depend on $\alpha _{0}$ and $q$ and corresponds to the current density in
the absence of the string. For this part one has%
\begin{equation}
\langle j^{z}\rangle _{c}^{(0)}=\frac{2m^{2}e}{\pi ^{2}L}\sum_{l=1}^{\infty }%
\frac{\sin (2\pi l\tilde{\beta})}{l}K_{2}(lLm)\ ,  \label{jzc00}
\end{equation}%
which does not depend on the radial coordinate $r$. Eq. (\ref{jzc00})
presents the current density in the Minkowski spacetime with the spatial
topology $R^{2}\times S^{1}$. It is a special case of a general formula
given in Ref. \cite{Bell10} for the topology $R^{p}\times \left(
S^{1}\right) ^{q}$ with arbitrary $p$ and $q$. As we see from (\ref{jzc}),
the axial current density vanishes for integer and half-integer values of $%
\tilde{\beta}$. In particular, this is the case for untwisted and twisted
fields in the absence of the magnetic flux enclosed by the string axis ($%
\Phi _{3}=0$).

The axial current density is finite on the string:
\begin{eqnarray}
\langle j^{z}\rangle _{c}|_{r=0} &=&\frac{4m^{2}e}{\pi ^{2}L}%
\sum_{l=1}^{\infty }\sin (2\pi l\tilde{\beta })\frac{K_{2}(mLl)}{l}\left[ %
\sideset{}{'}{\sum}_{k=0}^{p}(-1)^{k}c_{k}\cos \left( 2\pi k\alpha
_{0}\right) \right.  \notag \\
&&\left. +\frac{q}{\pi }\int_{0}^{\infty }dv\frac{h(q,\alpha _{0},2v)\sinh v%
}{\cosh (2qv)-\cos (q\pi )}\right] \ .  \label{jzc0}
\end{eqnarray}%
For the special case with (\ref{gammmaSp}) the integral term in (\ref{jzc})
vanishes and for the current density we get
\begin{equation}
\langle j^{z}\rangle _{c}=\frac{2m^{2}e}{\pi ^{2}L}\sum_{l=1}^{\infty }l\sin
(2\pi l\tilde{\beta })\sum_{k=0}^{q-1}c_{k}\cos \left( \pi k\frac{2n+1}{q}%
\right) \frac{K_{2}(mL\sqrt{l^{2}+\rho _{k}^{2}})}{l^{2}+\rho _{k}^{2}}.
\label{jzcSp}
\end{equation}%
For a massless field this formula simplifies to%
\begin{equation}
\langle j^{z}\rangle _{c}=\frac{4e}{\pi ^{2}L^{3}}\sum_{l=1}^{\infty }l\sin
(2\pi l\tilde{\beta })\sum_{k=0}^{q-1}\cos \left( \pi k\frac{2n+1}{q}\right)
\frac{c_{k}}{\left( l^{2}+\rho _{k}^{2}\right) ^{2}}.  \label{jzcSpm0}
\end{equation}

\begin{figure}[tbph]
\begin{center}
\epsfig{figure=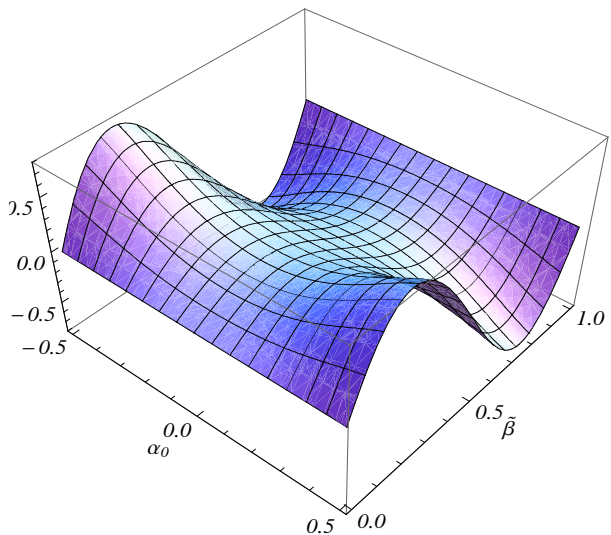,width=8.cm,height=7.cm}
\end{center}
\caption{Axial current density, $L^{3}\langle j^{z}\rangle _{c}/e$, as a
function of $\protect\alpha _{0}$ and $\tilde{\protect\beta}$ in the
geometry of a cosmic string with $q=2.5$ and for $r/L=0.25$.}
\label{fig4}
\end{figure}

For a massless field the general formula (\ref{jzc}) reduces to
\begin{eqnarray}
\langle j^{z}\rangle _{c} &=&\frac{8e}{\pi ^{2}L^{3}}\left[ %
\sideset{}{'}{\sum}_{k=0}^{p}(-1)^{k}c_{k}\cos \left( 2\pi k\alpha
_{0}\right) S(\tilde{\beta},\rho _{k})\right.   \notag \\
&&\left. +\frac{q}{\pi }\int_{0}^{\infty }dv\frac{\sinh vh(q,\alpha _{0},2v)%
}{\cosh (2qv)-\cos (q\pi )}S(\tilde{\beta},\eta (v))\right] ,  \label{jzcm0}
\end{eqnarray}%
where%
\begin{equation}
S(\tilde{\beta},x)=\sum_{l=1}^{\infty }\frac{l\sin (2\pi l\tilde{\beta})}{%
\left( l^{2}+x^{2}\right) ^{3}}.  \label{Sbet}
\end{equation}%
The function in (\ref{Sbet}) can be written as%
\begin{equation}
S(\tilde{\beta},x)=-\frac{1}{4x}\partial _{x}\sum_{l=1}^{\infty }\frac{l\sin
(2\pi l\tilde{\beta})}{\left( l^{2}+x^{2}\right) ^{2}},  \label{Sbet1}
\end{equation}%
and the expression for the series in the right-hand side can be found in
\cite{Prud86} (with the sign missprint corrected). In this way one gets:%
\begin{eqnarray}
S(\tilde{\beta},x) &=&\frac{\pi ^{2}}{16x^{3}\sinh y}\left\{ -\frac{\sinh (2%
\tilde{\beta}y)}{\sinh y}\left( 2y\coth y+1\right) \right.   \notag \\
&&\left. +2\tilde{\beta}\left[ 2y\frac{\cosh (2\tilde{\beta}y)}{\sinh y}+2%
\tilde{\beta}y\sinh [(1-2\tilde{\beta})y]+\cosh [(1-2\tilde{\beta})y]\right]
\right\} ,  \label{Sbet2}
\end{eqnarray}%
with $0\leqslant \tilde{\beta}\leqslant 1$ and $y=\pi x$. For $r\ll L$, the
leading term in the axial current is given by (\ref{jzcm0}) with the
replacement $S(\tilde{\beta},x)\rightarrow S(\tilde{\beta},0)$, where%
\begin{equation}
S(\tilde{\beta},0)=\frac{\pi ^{5}}{45}\tilde{\beta}(1-\tilde{\beta})(1-2%
\tilde{\beta})(1+3\tilde{\beta}-3\tilde{\beta}^{2}).  \label{Sbet0}
\end{equation}%
At large distances from the string the axial current density, given by (\ref%
{jzcm0}), is dominated by the $k=0$ term. In order to estimate the
contribution of the remaining part we note that for $0<\tilde{\beta}<1/2$
and assuming $\tilde{\beta}x\gg 1$ one has the asymptotic expression $S(%
\tilde{\beta},x)\approx \pi ^{3}(\tilde{\beta}/x)^{2}e^{-2\pi \tilde{\beta}%
x}/4$. Hence, this contribution is suppressed by the factor $\exp [-4\pi
\tilde{\beta}r\sin (\pi /q)/L]$ for $q\geqslant 2$ and by the factor $\exp
(-4\pi \tilde{\beta}r/L)$ for $q<2$. The asymptotic behavior for $1/2<\tilde{%
\beta}<1$ is obtained by using the property that the axial current density
changes the sign under the replacement $\tilde{\beta}\rightarrow 1-\tilde{%
\beta}$. In fig. \ref{fig4} we plot $L^{3}\langle j^{z}\rangle _{c}/e$ as a
function of $\alpha _{0}$ and $\tilde{\beta}$ in the geometry of a cosmic
string with the parameter $q=2.5$ and for $r/L=0.25$.

\section{Conclusion}

\label{conc}

In this paper we have investigated the influence of the non-trivial spatial
topology on the VEV of the fermionic current densities. The combined effects
of two types of topology change are considered. The first one is due to the
planar angle deficit induced by the cosmic string and the second one is
induced by the compactification of the string along its axis. Along the
compactified dimension we have considered a general quasiperiodicity
condition with an arbitrary phase $\beta $. As special cases, it includes
periodicity and antiperiodicity conditions corresponding to untwisted and
twisted fields respectively. As we could observe, all the new contributions
to the current induced by the compactification depend crucially on the
parameter $\beta $. In addition, we have assumed the presence of a constant
gauge field with nonzero azimuthal and axial components. They correspond to
a magnetic flux along the string axes and to a magnetic flux enclosed by the
compact dimension.

For the evaluation of the VEV for the current density we have used the
direct summation method over a complete set of modes. In this method a
complete set of fermionic mode functions are employed. The complete set of
positive- and negative-energy mode functions is constructed in section \ref%
{sec2} and they are given by the expressions (\ref{psi+n}) and (\ref{psi-}).
Given these functions, the VEV of the current density is presented in the
form \ of the mode sum (\ref{current}). We have shown that the charge
density and the radial component of the current density vanish. By making
use of the Abel-Plana-type summation formula (\ref{sumform}), the azimuthal
current density is decomposed into two parts. The first one corresponds to
the current density for the geometry of a straight cosmic string and the
second one is induced by the compactification of the string axis. For the
first part we have provided a closed form (\ref{jazimu}) for a massive
fermionic field valid for general value of the planar angle deficit. It
includes various special cases previously discussed in the literature. A
simple expression, (\ref{jazimu1}), is obtained for a special value (\ref%
{gammmaSp}) for the parameter $\alpha _{0}$ characterizing the magnetic flux
along the axis of the string and $q$ being an integer number.

The part in the azimuthal current density induced by the compactification of
the string axis is given by the formula (\ref{jc-4b}). This part is an odd
periodic function of the magnetic flux along the string axis with the period
equal to the flux quantum and it is an even function of the parameter $%
\tilde{\beta}$, defined by (\ref{bett}), with the period 1. Unlike to the
part $\langle j^{\phi }\rangle _{s}$, which diverges on the string as $r^{-4}
$, the topological part of the azimuthal current density is finite on the
string (see (\ref{jcNearStr})). For a massless field, the topological part
is further simplified and the total current density is given by the
expression (\ref{jcm0}). Near the string the total current density is
dominated by the part $\langle j^{\phi }\rangle _{s}$. For a massless field,
at distances from the string, larger than the length of the
compactification, the behavior of the azimuthal current density depends
crucially on whether $\tilde{\beta}=0$ or not. For $\tilde{\beta}\neq 0$ the
compactification of the string along its axis leads to the suppression of
the current density, whereas for $\tilde{\beta}=0$ the current density is
increased by the compactification: $\langle j^{\phi }\rangle /\langle
j^{\phi }\rangle _{s}\sim r/L$.

The compactification of the cosmic string axis, in combination with the
quasiperiodicity condition (\ref{Period}) and with the component of the
gauge field along the string axis, leads to the nonzero VEV of the axial
current density. The phases in the periodicity conditions and the axial
component of the gauge field are related to each other through a gauge
transformation and the physical results depend on the combination (\ref{bett}%
). The VEV of the axial current density is given by the expression (\ref{jzc}%
). This VEV has a purely topological origin and vanishes in the geometry of
a straight cosmic string. Of course, the latter is a direct consequence of
the problem symmetry. The axial current density is a periodic function of
the magnetic flux along the string axis with the period equal to the flux
quantum. It is an odd periodic function of the parameter $\tilde{\beta}$
with the period 1. The axial current density vanishes for integer and
half-integer values of $\tilde{\beta}$. In particular, this is the case for
untwisted and twisted fields in the absence of the magnetic flux enclosed by
the string axis. The axial current density is finite on the string's axis.
At large distances from the string, it tends to a limiting value which
corresponds to the current density in the Minkowski spacetime with the
spatial topology $R^{2}\times S^{1}$. For a massless field, the summation
over $l$ in (\ref{jzc}) is done explicitly and the expression for the axial
current density takes simpler form (\ref{jzcm0}).

\section*{Acknowledgments}

ERBM thanks Conselho Nacional de Desenvolvimento Cient\'{\i}fico e Tecnol%
\'{o}gico (CNPq) for partial financial support.


\begin{thebibliography}{99}
\bibitem{Most97} E. Elizalde, S.D. Odintsov, A. Romeo, A.A. Bytsenko and S.
Zerbini, \textit{Zeta regularization techniques with applications} (World
Scientific, Singapore, 1994); V.M. Mostepanenko and N.N. Trunov, \textit{The
Casimir Effect and Its Applications} (Clarendon, Oxford, 1997); K.A. Milton,
\textit{The Casimir Effect: Physical Manifestation of Zero-Point Energy}
(World Scientific, Singapore, 2002); M. Bordag, G.L. Klimchitskaya, U.
Mohideen, and V.M. Mostepanenko, \textit{Advances in the Casimir Effect}
(Oxford University Press, Oxford, 2009); \textit{Lecture Notes in Physics:
Casimir Physics}, Vol. 834, edited by D. Dalvit, P. Milonni, D. Roberts, and
F. da Rosa (Springer, Berlin, 2011).

\bibitem{Eliz01} E. Elizalde, Phys. Lett. B \textbf{516}, 143 (2001); C.L.
Gardner, Phys. Lett. B \textbf{524}, 21 (2002); K.A. Milton, Grav. Cosmol.
\textbf{9}, 66 (2003); A.A. Saharian, Phys. Rev. D \textbf{70}, 064026
(2004); E. Elizalde, J. Phys. A \textbf{39}, 6299 (2006); A.A. Saharian,
Phys. Rev. D \textbf{74}, 124009 (2006); B. Green and J. Levin, J. High
Energy Phys. \textbf{11}, 096 (2007); P. Burikham, A. Chatrabhuti, P.
Patcharamaneepakorn, and K. Pimsamarn, J. High Energy Phys. \textbf{07}, 013
(2008); A.R. Zhitnitsky, Phys. Rev. D \textbf{86}, 045026 (2012).

\bibitem{Mello12} E.R. Bezerra de Mello and A.A. Saharian, Class. Quantum
Grav. \textbf{29} 035006 (21012).

\bibitem{V-S} T.W.B. Kibble, Phys. Rep. \textbf{67}, 183 (1980); A. Vilenkin
and E.P.S. Shellard, \textit{Cosmic Strings and Other Topological Defects}
(Cambridge University Press, Cambridge, England, 1994).

\bibitem{Berezinski} V. Berezinski, B. Hnatyk and A. Vilenkin, Phys. Rev. D
\textbf{64}, 043004 (2001).

\bibitem{Damour} T. Damour and A. Vilenkin, Phys. Rev. Lett. \textbf{85},
3761 (2000).

\bibitem{Bhattacharjee} P. Bhattacharjee and G. Sigl, Phys. Rep. \textbf{327}%
, 109 (2000).

\bibitem{Sarangi} S. Sarangi and S. -H. Henry Tye, Phys. Lett. B \textbf{536}%
, 185 (2002).

\bibitem{Copeland} E.J. Copeland, R.C. Myers, and J. Polchinski, J. High
Energy Phys. \textbf{06}, 013 (2004).

\bibitem{Dvali} G. Dvali and A. Vilenkin, J. Cosmol. Astropart. Phys.
\textbf{03}, 010 (2004).

\bibitem{Hell86} T.M. Helliwell and D.A. Konkowski, Phys. Rev. D \textbf{34}
, 1918 (1986).

\bibitem{Line87} B. Linet, Phys. Rev. D \textbf{35}, 536 (1987).

\bibitem{Frol87} V.P. Frolov and E.M. Serebriany, Phys. Rev. D \textbf{35},
3779 (1987).

\bibitem{Dowk87} J.S. Dowker, Phys. Rev. D \textbf{36}, 3095 (1987); J.S.
Dowker, Phys. Rev. D \textbf{36}, 3742 (1987).

\bibitem{Davi88} P.C.W. Davies and V. Sahni, Class. Quantum Grav. \textbf{5}
, 1 (1988).

\bibitem{Smit89} A.G. Smith, in \textit{The Formation and Evolution of
Cosmic Strings}, Proceedings of the Cambridge Workshop, Cambridge, England,
1989, edited by G.W. Gibbons, S.W. Hawking, and T. Vachaspati (Cambridge
University Press, Cambridge, England, 1990).

\bibitem{Alle90} B. Allen and A.C. Ottewill, Phys. Rev. D \textbf{42}, 2669
(1990); B. Allen, J.G. Mc Laughlin, and A.C. Ottewill, Phys. Rev. D \textbf{%
45}, 4486 (1992); B. Allen, B.S. Kay, and A.C. Ottewill, Phys. Rev. D
\textbf{53}, 6829 (1996).

\bibitem{Sour92} T. Souradeep and V. Sahni, Phys. Rev. D \textbf{46}, 1616
(1992).

\bibitem{Shir92} K. Shiraishi and S. Hirenzaki, Class. Quantum Grav. \textbf{%
9}, 2277 (1992).

\bibitem{Beze94} V.B. Bezerra and E.R. Bezerra de Mello, Class. Quantum
Grav. \textbf{11}, 457 (1994); E.R. Bezerra de Mello, Class. Quantum Grav.
\textbf{11}, 1415 (1994).

\bibitem{Cogn94} G. Cognola, K. Kirsten, and L. Vanzo, Phys. Rev. D \textbf{%
49}, 1029 (1994).

\bibitem{More95} E.S. Moreira Jnr, Nucl. Phys. B \textbf{451}, 365 (1995).

\bibitem{Iell97} D. Iellici, Class. Quantum Grav. \textbf{14}, 3287 (1997).

\bibitem{Khus99} N.R. Khusnutdinov and M. Bordag, Phys. Rev. D \textbf{59},
064017 (1999).

\bibitem{BezeKh06} V.B. Bezerra and N.R. Khusnutdinov, Class. Quantum Grav.
\textbf{23}, 3449 (2006).

\bibitem{charged} J.S. Dowker, Phys. Rev. D \textbf{36}, 3742 (1987).

\bibitem{charged1} M.E.X. Guimar\~{a}es and B. Linet, Commun. Math. Phys.
\textbf{165}, 297 (1994).

\bibitem{charged3} J. Spinelly and E.R. Bezerra de Mello, Class. Quantum
Grav. \textbf{20} 874, (2003); J. Spinelly and E.R. Bezerra de Mello, Int.
J. Mod. Phys. A, \textbf{17}, 4375 (2002).

\bibitem{Spin} J. Spinelly and E.R. Bezerra de Mello, Int. J. Mod. Phys. D
\textbf{13}, 607 (2004); J. Spinelly and E.R. Bezerra de Mello, Nucl Phys. B
(Proc. Suppl.) \textbf{127}, 77 (2004).

\bibitem{Spin1} J. Spinelly and E. R. Bezerra de Mello, JHEP \textbf{09},
005 (2008).

\bibitem{Dou} M.R. Douglas and S. Kachru, Rev. Mod. Phys. \textbf{79}, 733
(2007).

\bibitem{Sira} L. Sriramkumar, Class. Quantum Grav. \textbf{18}, 1015 (2001).

\bibitem{Yu} Yu.A. Sitenko and N.D. Vlasii, Class. Quantum Grav. \textbf{26}%
, 195009 (2009).

\bibitem{Mello10} E.R. Bezerra de Mello, Class. Quantum Grav. \textbf{27},
095017 (2010).

\bibitem{Saha10} E.R. Bezerra de Mello, V.B. Bezerra, A.A. Saharian, and
V.M. Bardeghyan, Phys. Rev. D \textbf{82}, 085033 (2010).

\bibitem{Brev95} I. Brevik and T. Toverud, Class. Quantum Grav. \textbf{12},
1229 (1995); E.R. Bezerra de Mello, V.B. Bezerra, A.A. Saharian, and A.S.
Tarloyan, Phys. Rev. D \textbf{74}, 025017 (2006); E.R. Bezerra de Mello,
V.B. Bezerra, and A.A. Saharian, Phys. Lett. B \textbf{645}, 245 (2007); E.
R. Bezerra de Mello, V. B. Bezerra, A. A. Saharian, and A. S. Tarloyan,
Phys. Rev. D \textbf{78}, 105007 (2008); G. Fucci and K. Kirsten, JHEP
\textbf{1103}, 016 (2011); E.R. Bezerra de Mello and A.A. Saharian, Class.
Quantum Grav. \textbf{28}, 145008 (2011); G. Fucci and K. Kirsten, J. Phys.
A \textbf{44}, 295403 (2011); E.R. Bezerra de Mello, A.A. Saharian, and
A.Kh. Grigoryan, J. Phys. A: Math. Theor. \textbf{45}, 374011 (2012).

\bibitem{Cast09} A.H. Castro Neto, F. Guinea, N.M.R. Peres, K.S. Novoselov,
and A.K. Geim, Rev. Mod. Phys. \textbf{81}, 109 (2009).

\bibitem{Sous89} P. de Sousa Gerbert and R. Jackiw, Commun. Math. Phys.
\textbf{124}, 229 (1989); P. de Sousa Gerbert, Phys. Rev. D \textbf{40},
1346 (1989); Yu.A. Sitenko, Ann. Phys. \textbf{282}, 167 (2000).

\bibitem{Bene00} C.G. Beneventano, M. De Francia, K. Kirsten, and E.M.
Santangelo, Phys. Rev. D \textbf{61}, 085019 (2000); M. De Francia and K.
Kirsten, Phys. Rev. D \textbf{64}, 065021 (2001).

\bibitem{Bell11} S. Bellucci, E.R. Bezerra de Mello, and A.A. Saharian,
Phys. Rev. D \textbf{83}, 085017 (2011); E.R. Bezerra de Mello, F. Moraes,
A.A. Saharian, Phys. Rev. D \textbf{85}, 045016 (2012).

\bibitem{Jack09} R. Jackiw, A.I. Milstein, S.-Y. Pi, and I.S. Terekhov,
Phys. Rev. B \textbf{80}, 033413 (2009); A.I. Milstein and I.S. Terekhov,
Phys. Rev. B \textbf{83}, 075420 (2011).

\bibitem{Grad} I.S. Gradshteyn and I.M. Ryzhik, \textit{Table of Integrals,
Series and Products} (Academic Press, New York, 1980).

\bibitem{Bell10} S. Bellucci, A.A. Saharian, and V.M. Bardeghyan, Phys. Rev.
D \textbf{82}, 065011 (2010).

\bibitem{Bell13} S. Bellucci, A.A. Saharian, and H.A. Nersisyan,
arXiv:1302.1688.

\bibitem{SahaBook} A.A. Saharian, \textit{The Generalized Abel-Plana Formula
with Applications to Bessel Functions and Casimir Effect} (Yerevan State
University Publishing House, Yerevan, 2008); Report No. ICTP/2007/082;
arXiv:0708.1187.

\bibitem{Prud86} A.P. Prudnikov, Yu.A. Brychkov, and O.I. Marichev, \textit{%
Integrals and Series} (Gordon and Breach, New York, 1986), Vol. 2.
\end{thebibliography}
\end{document}